\documentclass{aip-cp}

\usepackage[numbers]{natbib}
\usepackage{rotating}
\usepackage{graphicx}
\usepackage{multirow}
\usepackage{subfigure}

\newcommand {\be} {\begin{equation}}
\newcommand {\ee} {\end{equation}}
\newcommand {\bea} {\begin{eqnarray}}
\newcommand {\eea} {\end{eqnarray}}

\begin{document}

\title{Models for SIMP dark matter and dark photon}

\author[aff1]{Hyun Min Lee\corref{cor1}}
\author[aff2]{Min-Seok Seo}


\affil[aff1]{Department of Physics, Chung-Ang University, 06974 Seoul, Korea}
\affil[aff2]{Center for Theoretical Physics of the Universe, Institute for Basic Science (IBS), 34051 Daejeon, Korea}
\corresp[cor1]{A talk given by H. M. Lee at PPC2015 conference, Deadwood, SD}

\maketitle

\begin{abstract}
We give a review on the SIMP paradigm and discuss a consistent model for SIMP dark mesons in the context of a dark QCD with flavor symmetry.  The $Z'$-portal interaction is introduced being compatible with stable dark mesons and is responsible for making the SIMP dark mesons remain in kinetic equilibrium with the SM during the freeze-out process. The SIMP parameter space of the $Z'$ gauge boson can be probed by future collider and direct detection experiments.  
\end{abstract}

 \section{Introduction}

WIMP paradigm is based on dark matter(DM) that is in chemical equilibrium with the SM particles through the $2\rightarrow 2$ annihilation of a dark matter pair into a pair of SM particles.
Once the $2\rightarrow 2$ annihilation rate is comparable to the Hubble rate, dark matter is decoupled from thermal plasma so that the DM relic density is determined in terms of weak-scale DM mass and weak interactions. But, it is known that a sizable DM annihilation into a pair of SM particles is not necessary for WIMP paradigm to work.  For instance, if dark matter can annihilate into a pair of mediator particles that are in kinetic equilibrium with thermal plasma, the WIMP mechanism still works. 
Therefore, other than the standard  
$2\rightarrow 2$ annihilation process involving the SM particles, any process changing the number of dark matter would be sufficient. Semi-annihilation occurring in models with $Z_N$ ($N>2$) discrete symmetry is such an example.  

Strongly Interacting Massive Particles (SIMP) have been recently considered as an alternative thermal dark matter to WIMP \cite{simp}. In this case, feeble interactions between dark matter and SM particles prevent dark matter from being in chemical equilibrium with the SM particles. Instead, the $3\rightarrow 2$ annihilation between dark matters is responsible for the thermal freeze-out, leading to a thermal dark matter of sub-GeV mass for an effective DM interaction of ${\cal O}(1)-{\cal O}(10)$.  
It is shown that SIMP paradigm can be realized in a dark QCD with a number of dark quarks where dark mesons are stable due to flavor symmetry and the Wess-Zumino-Witten(WZW) term is responsible for the effective 5-point interactions required for SIMP annihilation \cite{simpmeson}. The WZW term is a topological invariant existing for $\pi_5(G/H)=Z$ when flavor symmetry $G$ is spontaneously broken into $H$ by a QCD-like condensation and it is proportional to the number of colors $N_c$ in the dark QCD. Therefore, for a parametrically large $N_c$, it is possible to enhance the $3\rightarrow 2$ annihilation rate for dark mesons while making the $2\rightarrow 2$ self-scattering of dark matter consistent with the bounds from Bullet cluster and spherical hall shape. On a flip side, a large self-interaction of SIMP dark matter can solve small-scale structure problems, such as core-cusp and too-big-too-fail problems. 

In order for SIMP to be in kinetic equilibrium with the SM, however, there is a need of the minimum rate of the scattering of dark matter off of SM particles. In this case, the $2\rightarrow 2$ annihilation rate, coming from the crossing symmetry of the DM scattering diagrams, should be subdominant, as compared to the $3\rightarrow 2$ annihilation rate. 
In this article, we first give a review on the SIMP paradigm and discuss a consistent model of SIMP dark matter in the context of a dark QCD with $Z'$-portal interaction \cite{zportal} where SIMP dark mesons can be in kinetic equilibrium with the SM leptons until the freeze-out.   
The SIMP models with a fundamental scalar field have been also studied in the presence of a  discrete symmetry of global \cite{simp,rosenfeld} or local \cite{discrete,z5} origin. The merit of having a local discrete symmetry is that the $Z'$ gauge boson is built-in as a mediator field \cite{discrete,z5}.

\section{SIMP paradigm}

The $3\rightarrow 2$ annihilation rate for dark matter is $\frac{n^2_{\rm DM} v^2_{\rm DM}}{(2E_{\rm DM})^3}\int d\Pi_2 |{\cal M}_{3\rightarrow 2}|^2\delta^4(p_f-p_i)\equiv n^2_{\rm DM}(\sigma_{3\rightarrow 2} v^2_{\rm DM})$ where $d\Pi_2$ is the phase space integral for two DMs in the final states, $|{\cal M}_{3\rightarrow 2}|^2$ is the squared scattering amplitude being proportional to $\frac{1}{m^2_{\rm DM}}$ for scalar DM and $v_{\rm DM}$ is the DM velocity. As compared to the WIMP case, 
the  SIMP annihilation rate has a suppression factor due to the number density of extra DM particle by $\frac{n_{\rm DM}}{m^3_{\rm DM}}$ where $n_{\rm DM}\sim (m_{\rm DM }T_F)^{3/2} e^{-m_{\rm DM}/T}$ is Boltzmann-suppressed in the non-relativistic limit at the freeze-out temperature $T_F$.  Thus, imposing the freeze-out condition $\Gamma_{3\rightarrow 2}=H(T_F)$ requires a lower freeze-out temperature, meaning a smaller DM mass, due to $x_F\equiv m_{\rm DM}/T_F\approx 20$ similarly to the WIMP case.  For $\langle\sigma v^2_{\rm DM}\rangle_{3\rightarrow 2} \equiv \frac{\alpha^3_{\rm eff}}{m^5_{\rm DM}}$ where $\alpha_{\rm eff}$ is the effective DM coupling,  the relic density condition for SIMP dark matter is given \cite{simp} by
\be
m_{\rm DM}=\alpha_{\rm eff} \left(\frac{\kappa^2}{x^4_F}\sqrt{\frac{90}{g_*\pi^2}} \, T^2_{\rm eq} M_P \right)^{1/3}\simeq \alpha_{\rm eff} \times (35\,{\rm MeV})  \label{relic}
\ee
 where $T_{\rm eq}$ is the temperature at matter-radiation equality given by $T_{\rm eq}=0.8\,{\rm eV}$,
 and $g_*=10.75$ for $1\,{\rm MeV}\lesssim T\lesssim 100\,{\rm MeV}$, and $\kappa=\frac{2\pi^2}{45}g_{*s} c=2.55$ for $c\equiv 0.63\, g_{*,{\rm eq}}/g_{*s,{\rm eq}}=0.54$. 
Therefore, for $\alpha_{\rm eff}=1-30$, a sub-GeV thermal dark matter is a natural outcome of the SIMP paradigm.

\section{SIMP dark mesons with WZW term}

Suppose that dark flavor symmetry $G$ is spontaneously broken to a subgroup $H$ due to a QCD-like condensation. Then, the effective action for dark mesons is described by $U= e^{2i\pi /F}$ with $\pi\equiv \pi^a T^a$ where $F$ is the dark meson decay constant and $T^a$ are the broken generators of flavor symmetry. When the topological condition on the coset space, $\pi_5(G/H)=Z$, is satisfied, there exists a nonzero Wess-Zumino-Witten(WZW) term in the effective action for dark mesons, in the following form \cite{wzw1,wzw2},
\be
{\cal L}_{WZW}= \frac{2N_c}{15\pi^2} \epsilon^{\mu\nu\rho\sigma}{\rm Tr}[\pi \partial_\mu\pi \partial_\nu\pi \partial_\rho\pi \partial_\sigma \pi]
\ee 
with $N_c$ being the number of colors in dark QCD.  For instance, in the case with $G/H=SU(N_f)\times SU(N_f)/SU(N_f)$ for $SU(N_c)$ dark QCD with $N_f$ flavors, the topological condition is fulfilled for $N_f\geq 3$.  
Consequently, the WZW term yields 5-point interactions between dark mesons. When the unbroken flavor symmetry is respected by dark quark masses, dark mesons have equal masses from the effective interaction, ${\cal L}_{\rm mass}=-\frac{1}{2}\Lambda^3{\rm Tr}[M_q(U+U^{-1})]$, where $M_q$ is the quark mass matrix.  The thermal-averaged $3\rightarrow 2$ annihilation cross section is computed with the WZW term as follows \cite{simpmeson},
\be
\langle \sigma v^2\rangle_{3\rightarrow 2}=\frac{5\sqrt{5} N^2_c m^5_\pi}{2\pi^5 F^{10}}\frac{t^2}{N^3_\pi} \left(\frac{T_F}{m_\pi}\right)^2
\ee 
where $N_\pi$ is the number of dark mesons given by $N_\pi=N^2_f-1$ and $t^2=\frac{4}{3}N_f (N^2_f-1)(N^2_f-4)$ in the case of the coset space $SU(N_f)\times SU(N_f)/SU(N_f)$ with $SU(N_c)$ dark QCD. On the other hand, the self-scattering cross section for dark mesons is given by $\sigma_{\rm self}=\frac{m^2_\pi}{32\pi F^4} \frac{a^2}{N^2_\pi}$ with $a^2=8 (N_f^2-1)(3N^4_f-2 N^2_f+6)/N^2_f$ in the same case \cite{simpmeson}.  Consequently, the effective DM coupling for the $3\rightarrow 2$ annihilation is given by $\alpha_{\rm eff}=10\Big(\frac{N_c}{3}\Big)^{2/3} \Big(\frac{ m_\pi/F}{5}\Big)^{10/3}\Big(\frac{t^2/N^3_\pi}{(t^2 N^3_\pi)|_{N_f=3}}\Big)^{1/3}\Big(\frac{20}{x_F}\Big)^{2/3}$. 
Then,  for sub-GeV meson masses, the SIMP mesons can satisfy both the relic density condition (\ref{relic}) and the perturbativity condition,  $m_\pi/F<2\pi$.
On the other hand,  the self-scattering cross section can be rewritten as $\sigma_{\rm self}=\frac{160}{m^2_\pi}\Big(\frac{m_\pi/F}{5}\Big)^4\Big(\frac{a^2/N^2_\pi}{(a^2/N^2_\pi)|_{N_f=3}}\Big)$.
Thus, for $N_c=3$, $N_f=3$, and $m_\pi/F=5$, we get $m_\pi\sim 350\,{\rm MeV}$ and $\sigma_{\rm self}/m_\pi\sim 3{\rm cm^2/g}$, which exceeds the bound from Bullet cluster but is sufficiently large to solve small-scale structure problems and explain the separation of the DM halo in Abell 3827 \cite{abell}. But, in order to satisfy the Bullet cluster bound, $\sigma_{\rm self}/m_\pi < 1{\rm cm^2/g}$, 
we need to take a smaller $m_\pi/F$, which implies $N_c>3$ for the fixed value of $\alpha_{\rm eff}$.

\section{Communication to dark mesons via $Z'$-portal}

SIMP paradigm is based on the assumption that dark matter remains in kinetic equilibrium with the SM during the freeze-out process. Instead, a dark temperature can be introduced such that dark radiation is subdominant to avoid BBN or CMB bounds, and it is necessary for a light thermal dark matter with $m_{\rm DM}<1\,{\rm MeV}$, not to affect the structure formation. For SIMP paradigm with 5-point DM interactions, dark matter mass lies between ${\cal O}(10)\,{\rm MeV}$ and sub-GeV, so we don't have to introduce a dark temperature.  Parametrizing the scattering cross section between SIMP and an SM fermion by $\langle\sigma v\rangle_{\rm scatt}=\frac{\delta^2_1}{m^2_{\rm DM}}$, the kinetic equilibrium condition leads to $\delta_1 \gtrsim 10^{-9}\alpha^{1/2}_{\rm eff}$, while the $2\rightarrow 2$ annihilation cross section occurring due to crossing symmetry is given by $\langle\sigma v\rangle_{\rm ann}=\frac{\delta^2_2}{m^2_{\rm DM}}$ and subdominant if $\delta_2\lesssim 2.4 \times 10^{-6}\alpha_{\rm eff}$.
In order to make an SM-singlet dark matter in kinetic equilibrium with the SM, we need to introduce a mediator field between dark matter and the SM, such as Higgs or $Z'$-portals. Since the SIMP freeze-out temperature is quite low as $T_F\simeq \frac{m_{\rm DM}}{20}\lesssim 50\,{\rm MeV}$ for $m_{\rm DM}\lesssim 1\,{\rm GeV}$, it is  only through scattering off of electron/positron, neutrinos and photon that SIMP can be made in kinetic equilibrium. In the case of Higgs-portal, the scattering cross section is suppressed by the electron Yukawa coupling so it is not sufficient for kinetic equilibrium \cite{rosenfeld,zportal,discrete}. 
On the other hand, the $Z'$-portal coupling to the electron is proportional to the electric charge so it is appropriate for making SIMP in kinetic equilibrium \cite{zportal,discrete}.  

We assume that dark quarks are vector-like under a local dark $U(1)$. 
Then, the WZW term is modified in the presence of the local $U(1)$ as follows \cite{wzw2},
\bea
S&=&S_0(D_\mu U, D_\mu U^{-1})+S_{WZW}(U,U^{-1})-eN_c \int d^4 x A'_\mu J^\mu \nonumber \\
&&+\frac{i g^2_D N_c}{24\pi^2} \int d^4x \epsilon^{\mu\nu\rho\sigma} \partial_\mu A^{\prime}_ \nu A'_\rho {\rm Tr}[Q^2_D \partial_\sigma U U^{-1}+Q^2_D U^{-1}\partial_\sigma U+Q_D U Q_D U^{-1} \partial_\sigma U U^{-1}]  \label{wzwp}
\eea
where $S_0$ is the chiral Lagrangian for dark mesons, $J^\mu$ is the Noether current for dark mesons, $Q_D$ is the dark charge operator, and the covariant derivative is defined as $D_\mu U=\partial_\mu U+ig_D [Q_D,U] A'_\mu$. As  a consequence, the coupling between dark meson $\pi^a$ and a pair of dark $Z'$, the last term in eq.~(\ref{wzwp}), is generated due to chiral anomalies, being proportional to ${\rm Tr}(Q^2_D T^a)$. When $Z'$ is massless or light enough, those anomaly couplings would make dark mesons decay in a very short time.
Even in the case of massive $Z'$, a gauge kinetic mixing between $Z'$ and the hypercharge gauge boson makes dark mesons decay into SM leptons again promptly.  
Therefore, we need to cancel the chiral anomalies by imposing ${\rm Tr}(Q^2_D T^a)=0$. 
For the coset space $SU(3)_f\times SU(3)_f/SU(3)_f$ with $SU(3)_c$ color group, we can cancel the chiral anomalies by taking the dark charge operator \cite{zportal} to be 
\be
Q_D=\left( \begin{array}{ccc} 1 & 0 & 0 \\  0 & -1 & 0  \\ 0 & 0 & -1 \end{array} \\ \right),
\ee
as compared to the SM case, $Q={\rm diag}(\frac{2}{3},-\frac{1}{3},-\frac{1}{3})$. Then, dark charged pions and kaons carry $\pm 2$ dark charges such that they can scatter off of the electron efficiently in the presence of a gauge kinetic mixing of the $Z'$ gauge field.  
On the other hand, the minimal $Z'$ coupling between dark mesons and $Z'$ could lead to the $2\rightarrow 2$ annihilation of dark mesons such as via $\pi\pi\rightarrow Z' Z'$ and $\pi\pi\rightarrow \pi Z'$ in the dark sector.  
However, we can make $Z'$ sufficiently heavy such that such a $2\rightarrow 2$ annihilation is sub-dominant. We also remark that from ${\cal L}_{\rm split}=-c\alpha_D \Lambda^4_c {\rm Tr}[Q_DU Q_DU^{-1}]$ with $\alpha_D\equiv \frac{g^2_D}{4\pi}$, $c$ being a constant of order one, and $\Lambda_c$ being the dark QCD scale, the non-universal dark charges lead to a mass splitting between neutral and charged dark mesons,  $\Delta m^2_\pi\sim \frac{\alpha_D \Lambda^4_c}{F^2}\sim \alpha_D F^2$, where use is made of $\Lambda_c\sim F$ in the last equality \cite{zportal}. Since the 5-point interactions involve both neutral and charged mesons, the mass splitting between dark mesons must be less than about $10\%$ in order for all mesons to exist for annihilation, requiring $\alpha_D\lesssim 0.01 (m_\pi/F)^2$.

In the presence of a gauge kinetic mixing, ${\cal L}_{\rm mix}=-\frac{\varepsilon}{2\cos\theta_W}\, F'_{\mu\nu} F^{\mu\nu}$,  the thermal-averaged scattering cross section between dark meson and charged lepton is given \cite{zportal,discrete} by
\be
\langle \sigma v\rangle_{{\rm scatt},l} \approx 96\pi \alpha\alpha_D \varepsilon^2\, \frac{m^2_\pi}{m^4_{Z'}} \left(\frac{T_F}{m_\pi} \right). 
\ee
Similarly, the scattering cross section between dark meson and the SM charged pion is $\langle\sigma v\rangle_{\rm scatt,\pi}\approx \langle \sigma v\rangle_{{\rm scatt},l}$.
 The kinetic equilibrium condition gives rise to the lower bound on the gauge kinetic mixing.
For instance, for $N_f=3$ and $\alpha_D=0.01$, we get  $\varepsilon\, \Big(\frac{m_\pi}{m_{Z'}}\Big)^2\gtrsim 10^{-8}$. The $2\rightarrow 2$ annihilation of dark mesons into leptons is also given \cite{zportal,discrete} by 
\be
\langle\sigma v\rangle_{{\rm ann},l} \approx 128\pi \alpha\alpha_D \varepsilon^2\, \frac{m^2_\pi}{(4m^2_\chi-m^2_{Z'})^2} \left(\frac{T_F}{m_\pi} \right). 
\ee
Therefore, the $2\rightarrow 2$ annihilation of dark mesons is subdominant if $\frac{ \varepsilon \,m^2_\pi}{|m^2_{Z'}-4m^2_\pi|}\lesssim 10^{-4}$. Combining the two SIMP conditions, the gauge kinetic mixing is bounded to $10^{-8}(m_{Z'}/m_\pi)^2 \lesssim\varepsilon\lesssim10^{-4} |m^2_{Z'}/m^2_\pi-4| $. The important message is that the SIMP condition leads to a lower bound on the gauge kinetic mixing. 

The $Z'$ searches at colliders give limits on $\varepsilon$ vs $m_{Z'}$, also depending on the dark gauge coupling when $Z'$ decays into a pair of dark matter.  $Z'$ searches with monophoton$+$dileptons at BaBar shows $\varepsilon \lesssim 6\times 10^{-4}$ for $0.01\,{\rm GeV}< m_{Z'}< 10.2\,{\rm GeV}$, but the limits on $\varepsilon$ become weaker by a factor $[\alpha_D/(\varepsilon^2 \alpha)]^{1/2}$ when $Z'$ decays invisibly into a pair of dark matter.  The current bound from the monophoton$+$MET at BaBar is at the level of $\varepsilon=10^{-3}$ below $m_{Z'}=8\,{\rm GeV}$ and it could be improved at Belle II. On the other hand, the beam dump experiments limit $\varepsilon$ at the level of $10^{-3}$ below $m_{Z'}=0.1\,{\rm GeV}$ and the limits could be improved in the planned SPS target experiment at CERN. In the case of heavier $Z'$,  the current bounds from the Higgs decay mode, $h\rightarrow Z Z'$ and resonance searches with dileptons at LHC, as well as EWPT, are at the level of $\varepsilon=10^{-2}$.  Then, the SIMP parameter space can be probed further by the future $Z'$ searches at colliders. 
The references and plots for various collider limits on the model can be found 
in our previous papers \cite{zportal,discrete}. 

\section{Outlook}

SIMP dark mesons with $Z'$-portal interaction can be a consistent thermal dark matter of sub-GeV mass. Direct detection experiments with superconducting detectors might probe the recoil energy of order $10\,{\rm eV}$ coming from electrons scattered off of SIMP dark matter after collision \cite{zurek,discrete}. However, the annihilation cross section of SIMP into leptons via $Z'$-portal is p-wave suppressed so it is too small to be bounded by cosmic ray observations at present \cite{discrete}. Observation of cosmic ray leptons below sub-GeV in the future might rule out the possibility of $Z'$-portal interaction and call for an alternative mediation mechanism.  
Furthermore, as SIMP dark mesons are on the verge of violating the perturbativity, it would be worthwhile to investigate a non-perturbative mechanism for enhancing the 5-point interactions, for instance, through the resonance of an extra singlet scalar in the SIMP model with a $Z_5$ discrete symmetry \cite{z5}.

\vspace{0.2cm}
\section{ACKNOWLEDGMENTS}
The work of HML is supported in part by Basic Science Research Program through the National Research Foundation of Korea (NRF) funded by the Ministry of Education, Science and Technology (2013R1A1A2007919). The work of MS is supported by IBS-R018-D1.



\end{document}